\documentclass[conference]{IEEEtran}
\IEEEoverridecommandlockouts
\usepackage{cite}
\usepackage{amsmath,amssymb,amsfonts}
\usepackage{algorithmic}
\usepackage{graphicx}
\usepackage{textcomp}
\usepackage{booktabs}
\usepackage{subfigure}
\usepackage{xcolor}
\usepackage{multirow}
\usepackage[super]{nth}
\def\BibTeX{{\rm B\kern-.05em{\sc i\kern-.025em b}\kern-.08em
    T\kern-.1667em\lower.7ex\hbox{E}\kern-.125emX}}
\begin{document}

\title{Target-centered Subject Transfer Framework for EEG Data Augmentation
\footnote{{\thanks{This work was supported by Institute of Information \& Communications Technology Planning \& Evaluation (IITP) grants funded by the Korea government (No. 2015-0-00185, No. 2017-0-00451)}
}}
}
\makeatletter
\newcommand{\linebreakand}{%
\end{@IEEEauthorhalign}
\hfill\mbox{}\par
\mbox{}\hfill\begin{@IEEEauthorhalign}
}
\makeatother

\author{

\IEEEauthorblockN{~~~~~~~Kang Yin}
\IEEEauthorblockA{\textit{~~~~~~~~~Dept. Artificial Intelligence} \\
\textit{~~~~~~~~~Korea University} \\
~~~~~~~~~Seoul, Republic of Korea \\
~~~~~~~~~charles\_kang@korea.ac.kr} \\ 

\and
\IEEEauthorblockN{~~~~~~~~~Byeong-Hoo Lee}
\IEEEauthorblockA{\textit{~~~~~~~~~Dept. Brain and Cognitive Engineering} \\
\textit{~~~~~~~~~Korea University} \\
~~~~~~~~~Seoul, Republic of Korea \\
~~~~~~~~~bh\_lee@korea.ac.kr} \\

\linebreakand
\IEEEauthorblockN{Byoung-Hee Kwon}
\IEEEauthorblockA{\textit{Dept. Brain and Cognitive Engineering} \\
	\textit{Korea University} \\
	Seoul, Republic of Korea \\
	bh\_kwon@korea.ac.kr} \\
\and
\IEEEauthorblockN{Jeong-Hyun Cho}
\IEEEauthorblockA{\textit{Dept. Brain and Cognitive Engineering} \\
\textit{Korea University} \\
Seoul, Republic of Korea \\
jh\_cho@korea.ac.kr}
}
\maketitle
\begin{abstract}
Data augmentation approaches are widely explored for the enhancement of decoding electroencephalogram signals. In subject-independent brain-computer interface system, domain adaption and generalization are utilized to shift source subjects' data distribution to match the target subject as an augmentation. However, previous works either introduce noises (e.g., by noise addition or generation with random noises) or modify target data, thus, cannot well depict the target data distribution and hinder further analysis. In this paper, we propose a target-centered subject transfer framework as a data augmentation approach. A subset of source data is first constructed to maximize the source-target relevance. Then, the generative model is applied to transfer the data to target domain. The proposed framework enriches the explainability of target domain by adding extra real data, instead of noises. It shows superior performance compared with other data augmentation methods. Extensive experiments are conducted to verify the effectiveness and robustness of our approach as a prosperous tool for further research. 
\end{abstract}

\begin{small}
\end{small}

\section{Introduction}
\label{sec:intro}
\iffalse
\begin{figure*}[tb]
	
	\centering    %子图居中
	\includegraphics[width=\textwidth]{./figures/figure1_icassp.pdf}
	\caption{Overview of the proposed framework. EEG data is first split by source subjects and target subject, and data relevance is maximized between the two. Given the optimized data, the Cycle-GAN based subject transfer is performed, and the pretained models are used to generate augmented data for downstream classification tasks.}
	\label{fig1} 
\end{figure*}
\fi

Motor imagery (MI) based Brain-Computer Interfaces (BCI) systems have experienced rapid developments since last century 90s~\cite{4}. The users are instructed to perform mental tasks, during which the electroencephalogram (EEG) signals are collected for the system to translate human's intention and makes corresponding response. It is generally believed that the successful decoding of users' intention heavily relies on their motivation and cognitive arousal for skill acquisition~\cite{6,9}, while distraction and drowsiness are harmful, thus results in poor signal qualities. Besides, individual uniqueness and specificity makes it difficult to conduct generalized analysis among subjects. Consequently, calibration process is usually inevitable before each online experiment, to fine-tune the pre-trained model to adapt a specific subject. This repeated process is tedious and consumes huge energy of subjects thus impairs their following MI performance.

It should be worth noticing that, unlike images or texts, EEG data may contain various noises originally, e.g. eye blinking, teeth clenching and flexing of both arms, thus honestly it's unwise to manually introduce more noises for augmentation. On one hand, researches are conducted to eliminate external perturbations~\cite{35,36}, while on the other hand, noises are added for generalization ability~\cite{37,38,39}, regardless of which, a reasonable balance ought to be struck between the two, and arbitrarily introducing noises goes against the aim of EEG data distillation and disentanglement, although experimental performance might get improved somehow. Besides, given a large amount of augmented data, it's not an optimal to learn a shared latent space of the augmented ones and the original ones~\cite{19,40,41}, as the newly included training data also transform the distribution of target subject in order for generalization. Consequently, the proportion of augmented data and raw one is usually very sensitive and required to be carefully searched.

Given the above observations, we argue that, theoretically, a practical DA approach in BCI should meet the following two conditions: (a) improve generalization ability without adding extra noises; (b) the data before and after augmentation should match with each other. It is encouraging that the basic SI BCI approach -- directly training with the whole dataset, actually doesn't introduce any noise, thus, the only problem lies in the distribution gap between a specific subject and the whole dataset. Inspired by remarkable works~\cite{20,3,12} that gain huge successes on image style transformation and regeneration in the field of computer vision, we assert that this kind of style transfer shows strong potentials also for EEG signals in BCI domain, with which the distribution difference is minimized between an individual and the all. 

Based on the above assumption, we propose a style transfer founded data augmentation framework. Different from a similar work~\cite{13,32,33,34} that modifies target subject's data distribution and create a co-domain for source and target data, the proposed one is designed to fit and perfect the domain for each target subject, i.e., the target domain's distribution remains relatively fixed and is gradually updated along with the arrival of transferred data from source domains. Substantial experiments have been conducted, and the proposed method shows superior performance on a very challenging multi-task motor imagery datasets. The visualization analysis on the generated data further proves the practicability and stability of our proposed method.

The main contributions of this paper are summarized as follows:

\begin{itemize}
	\item The proposed framework is the first to emphasize the importance of using real data for augmentation, and minimize the intersubject variability by distribution transfer to target domain.
	\item The source-target relevance maximization strategy optimizes the amount and quality of source data to ensure effectiveness and stability of the transfer.
\end{itemize}

\section{Methods}
\label{sec:method}
\subsection{Dataset description and preprocessing}

The OpenBMI dataset~\cite{21} is used in this work. It contains 200 trials for each of 54 subjects and for each of left and right hand motor imagery (MI) task. The MI task lasts for 4 s for each trial, and EEG signals were recorded with a sampling rate of 1,000 Hz and collected with 62 Ag/AgCl electrodes.

The EEG data were band-pass filtered between 8 and 30 Hz with a 5th order Butterworth digital filter, and is gone through z-score normalization. Besides, we reduce the original 62 channels to 8 for simplicity, where only [F3, C3, P3, Cz, Pz, F4, C4, P4] are reserved.

\subsection{Self-adjustable generative learning}
\subsubsection{Problem formulation}

\iffalse
\begin{table}[b]
	\caption{Notations in the method}
	\resizebox{\linewidth}{!}{
		\begin{tabular}{cl}
			\toprule
			\textbf{Variable} & \textbf{Description} \\
			\midrule
			$x_a, x_a^\prime$ & data of source subjects before and after subject transfer\\
			$x_b, x_b^\prime$ & data of target subject before and after subject transfer\\			
			$A, B$ & Domain of source subjects and target subject\\			
			$G_A, D_A$ & generator and discriminator in the domain of source data\\			
			$G_B, D_B$ & generator and discriminator in the domain of target data\\		
			\bottomrule
	\end{tabular}}
	\label{tab1}
\end{table}
\fi

Given a motor imagery dataset $X=[x^1, x^2, \dots, x^m]^T$, which contains EEG signals from $m$ subjects, and each $x^i$, where $i \in \{1,2,\cdots, m\}$, is a $t\times c\times p$ matrix, where $t$, $c$, $p$ denote the number of trials, channels and sample points, respectively. Trials from different subjects are concatenated by row in matrix $X$. Each subject is in turn taken as the target subject. Let $k$ be the current target subject number, accordingly, $(x^k)^{\complement}$ are taken as source subjects. For simplicity (Table~\ref{tab1}), let $a$ and $b$ denote source subjects and target subject. Data after subject transfer are marked by $\tilde{x}$.

\subsubsection{Source-target relevance maximization}

Intuitively, it is claimed that not all data is suitable for transfer, where outliers might exist, and it is of vital importance to select the most relative data for a target subject. In this section, we propose a two-step data filtering approach that optimize source data quality in order for better transfer performance in Sect.~\ref{transfer}. 

To be specific, principal component analysis (PCA) is firstly applied to reduce the dimension of original data, where data for each subject after dimension reduction is denoted by $y^i$. Then, we rule out inner-subject outliers, where the anomaly removal happened for each subject $y^i$. Finally, the cluster head $y_c$ of target data is calculated and the intra-subject optimal data selection is performed.

To perform PCA, we first calculate the mean $\overline{X}$ over channels:
\begin{equation}
	\overline{X}=\frac{\sum_{j=1}^{C} x_{:, j, :}}{C},
	\label{eq1}
\end{equation}
where $C$ is the number of channels.
\iffalse
Construct the covariance matrix $S$:
\begin{equation}
	S=\frac{1}{N}\overline{X}^TH\overline{X},
	\label{eq2}
\end{equation}
where $H$ is a $N$-dimensional centering matrix and $N$ is the number of samples. Based on singular value decomposition (SVD):
\begin{equation}
	H\overline{X}=U\Sigma V^T.
	\label{eq3}
\end{equation}
Considering both Eq.~\ref{eq2} and Eq.~\ref{eq3}, the covariance matrix could be rewritten as:
\begin{equation}
	S=\frac{1}{N}V\Sigma^2 V^T,
	\label{eq4}
\end{equation}
where $\Sigma$ and $V$ is the eigenvalue and eigenvector, respectively. Therefore, the data $Y$ after dimension reduction is formulated as:
\fi
By applying singular value decomposition (SVD):
\begin{equation}
	Y=H\overline{X}V,
	\label{eq5}
\end{equation}
where $H$ and $V$ are centering matrix and eigenvector matrix, respectively. $Y$ represents the data after dimension reduction.

After that, for each subject $y^i$, the cluster center $y^i_c$ is calculated:
\begin{equation}
	y_c^i=\frac{\sum_{j=1}^{t} y_{i, :}}{t},
\end{equation}
and the L1-distance $d_j$ is calculated to measure the relevance between a certain trial $y_j^i$ and its mean.\iffalse: 
\begin{equation}
	d_j^i=||y_j^i, y_c^i||.
\end{equation}
\fi 

For all subjects except the target one, a threshold $\beta_1$ is set to keep a proportion of data in the order of distance to the center point. Similarly, the distance between all remaining data and the target center point $y_c^{target}$ is calculated, and a threshold $\beta_2$ here will filter out those irrelevant trials which are unsuitable for transfer.

\subsubsection{Target-centered subject transfer}
\label{transfer}

Following the idea of Cycle-GAN in~\cite{20}, we aim to make source data $x_a$ and target data $x_b$ learn an approach to be capable of transferring to each other's style, by assuming them belong to two different domains. The thinking behind this assumption is intuitive: subjects vary a lot in many aspects, e.g., imagining ability, health condition and level of concentration, which are unique in individual level, thus every single subject should be accounted as a domain. We simply view source subjects as a whole domain in order to reduce the complexity of our method. After the subject transfer, data in source domain (various subjects) should obey the distribution in target domain (a single target subject), where  exterior differences caused by subjects no longer exists. 
\iffalse
\begin{figure}[tb]
	\centering  %居中
	%\subfigure[Classic Cycle GAN transfer]{
		
		\includegraphics[width=0.8\linewidth]{./figures/figure2_a.pdf}
		\caption{Illustration of target-centered subject transfer.}
		\label{fig2} 
	\end{figure}
\fi
	The subject transfer learning proceeds in a bi-directional manner, where two GANs are trained simultaneously, in which $(G_A, D_A)$ is used to transfer source data $x_a$ to target domain, and $(G_B, D_B)$  transfers target data $x_b$ to source domain. Accordingly, discriminators attempt to tell the real target (source) domain from source (target) domain, and generators gradually learn to transfer between domain by adversarially training with discriminators. As GAN is notorious for the unstable training process, here we follow previous works of Wasserstein GAN~\cite{26}, and combine it with traditional Cycle-GAN for better training, thus, the updated GAN loss is formulated as:
	\begin{equation}
		\begin{aligned}
			&\mathcal{L}_{\rm gan}(G, D, A, B)\\
			&={\mathbb E}_{x_b\sim {\mathbb P}_{B}}[\log D_A(G_B(x_b))]-{\mathbb E}_{x_a\sim {\mathbb P}_{A}}[\log D_A(x_a)]\\ 
			&+ {\mathbb E}_{x_a\sim {\mathbb P}_{A}}[\log D_B(G_A(x_a))]-{\mathbb E}_{x_b\sim {\mathbb P}_{B}}[\log D_B(x_b)]\\ 
			&+ \lambda{\mathbb E}_{\hat{x}\sim {\mathbb P}_{\hat{x}} }[(||\nabla_{\hat{x}}D_A(\hat{x})||_2-1)^2+ (||\nabla_{\hat{x}}D_B(\hat{x})||_2-1)^2],
		\end{aligned}
	\end{equation}
	where $\lambda$ is a hyperparameter, and the distribution ${\mathbb P}_{\hat{x}}$ obeys the uniform sampling rule along the line of pair-wise points sampled from ${\mathbb P}_{A}$ and ${\mathbb P}_{B}$.
	
	With the constraint of cycle consistence, the transferred data will be able to recover back to its original style by passing through the generator in corresponding domain, where the recovered data and the original one are asked to be as close as possible, therefore, the cycle consistence loss is written as follows:
	\begin{equation}
		\begin{aligned}
			\mathcal{L}_{\rm cyc}(G, A, B)&={\mathbb E}_{x_a\sim {\mathbb P}_{A}}[||G_B(G_A(x_a))-x_a||_1] \\
			&+{\mathbb E}_{x_b\sim {\mathbb P}_{B}}[||G_A(G_B(x_b))-x_b||_1].\\
		\end{aligned}
	\end{equation}
	\iffalse 
	To summarize, the overall loss function of the proposed subject transfer method is written as follows:
	\begin{equation}
		\begin{aligned}
			\mathcal{L}&=\mathcal{L}_{\rm gan}+\alpha\mathcal{L}_{\rm cyc}.\\
		\end{aligned}
	\end{equation}	
	\fi
	\section{Experiments and discussion}
	\label{sec:result}
	
	The experiment consists generative learning part, where augmented data are generated, and the downstream part for motor imagery classification. In the generative learning part, we implement  comparative method including both noise addition (noise, multiple and flip) and augmentation approaches in previous works ($\beta$-VAE~\cite{22}, DCGAN~\cite{23} and EEG-GAN~\cite{24}). In the classification part, EEGNet~\cite{25} is chosen as the backbone, and we following the implementations and hyperparameter settings in the original paper. Three RTX 3080 Ti GPUs are used in this experiment. 
	
	Three typical noise addition approaches are applied here. Let $d$ denotes the raw data, and the first approach is to add noise from a uniform distribution. The artificial data is generated by: $d_{noise}= d + \frac{u\cdot \text{std}(d)}{\gamma}$, where $u$ is sample from U(-1,1), and std($\cdot$) calculates the variance. The hyperparameter $\gamma$ is for scaling. Similarly, by multiplying and flip the raw data, we generate $d_{multi}=(1+\gamma)\times d$ and $d_{flip}= -d + \min(d)$.
	
	\subsection{Classification}
	\begin{table}[tb]
		\caption{Comparison of average classification accuracy (\%) for different methodologies.}
		\resizebox{\linewidth}{!}{
			\begin{tabular}{llll}
				\toprule
				Methodology & Mean (SD)  &Median&Range (Max–Min)\\
				\midrule
				Baseline &79.95 (9.91)&79.30&38.93 (97.28-58.35)\\
				\midrule
				%			\textbf{Noise addition} &&&\\
				Noise &80.82 (9.77)&80.45&38.36 (98.52-60.16)\\
				Multiple &80.07 (9.59)&79.02&36.76 (97.61-60.85)\\
				Flip &80.11 (9.56)&79.34& 35.56 (97.78-62.22)\\ 
				\midrule
				%			\textbf{Generative models} &&&\\
				$\beta$-VAE~\cite{22} &80.40 (9.03)&79.75&36.74 (94.87-58.13)\\
				DCGAN~\cite{23}&81.85 (9.85)&81.40& 39.84 (98.52-58.68)\\
				EEG-GAN~\cite{24}&81.03 (9.97)&\textbf{81.85}&41.73 (98.77-57.04)\\
				\midrule
				Cycle-GAN (Ours) &\textbf{82.34 (8.98)}&81.52&\textbf{34.05 (99.08-65.03)}\\
				\bottomrule
		\end{tabular}}
		\label{tab1}
	\end{table}
	Table~\ref{tab1} shows the classification results for different augmentation methods. Performance improvement is observed for all. Particularly, in noise addition approaches, multiple and flip operations could not effectively improve the accuracy, while by adding uniform distribution noises, 0.87\% classification accuracy increase is achieved. For generative models, we witness a clear superiority over than by simply adding noise. Also, the GAN based models obtain obviously better performance than VAE, which might be due to the poor reconstruction ability caused by its inherent architecture deficiency, where more comparison studies are provided in Sect.~\ref{diss}. Most importantly, our proposed method outperforms the other two GAN methods which are specially designed for EEG data, which shows the significance of using real signal data for generation, given that the other two both use random noises as the model input.
	
	\begin{figure}[tb]
		\centering  %居中

			\begin{minipage}{0.15\linewidth}
				\subfigure[Raw]{
				\centering
				\includegraphics[scale=0.4]{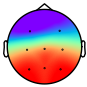}}
			\end{minipage}
			\begin{minipage}{0.15\linewidth}
				\subfigure[x10]{
				\centering    %子图居中
				\includegraphics[scale=0.43]{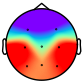}
			}
			\end{minipage}
			\begin{minipage}{0.15\linewidth}
				\subfigure[x20]{
				\centering    %子图居中
				\includegraphics[scale=0.44]{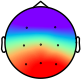}
					}
			\end{minipage}
			\begin{minipage}{0.15\linewidth}
					\subfigure[x30]{
				\centering    %子图居中
				\includegraphics[scale=0.46]{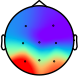}
					}
			\end{minipage}
				\begin{minipage}{0.15\linewidth}
				\subfigure[x40]{
				\centering    %子图居中
				\includegraphics[scale=0.45]{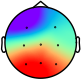}
					}
			\end{minipage}
				\begin{minipage}{0.15\linewidth}
					\subfigure[x50]{
				\centering    %子图居中
				\includegraphics[scale=0.43]{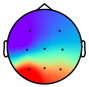}
					}
			\end{minipage}
	
		\caption{Topographic visualization of the power spectral density.}
		\label{erds} 
	\end{figure}
		\begin{figure}[b]
		\centering  %居中		
		\begin{minipage}{0.3\linewidth}
			\subfigure[Raw data]{
				\centering
				\includegraphics[scale=0.3]{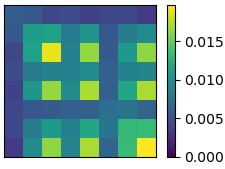}}
		\end{minipage}
		\begin{minipage}{0.3\linewidth}
			\subfigure[$\beta$-VAE]{
				\centering    %子图居中
				\includegraphics[scale=0.3]{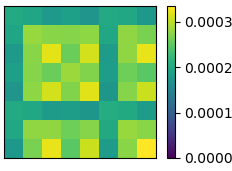}
			}
		\end{minipage}
		\begin{minipage}{0.3\linewidth}
			\subfigure[DCGAN]{
				\centering    %子图居中
				\includegraphics[scale=0.3]{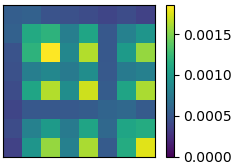}
			}
		\end{minipage}
		\begin{minipage}{0.3\linewidth}
			\subfigure[EEG-GAN]{
				\centering    %子图居中
				\includegraphics[scale=0.3]{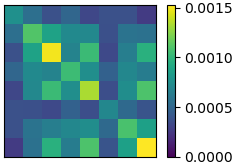}
			}
		\end{minipage}
		\begin{minipage}{0.3\linewidth}
			\subfigure[Ours]{
				\centering    %子图居中
				\includegraphics[scale=0.3]{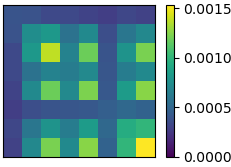}
			}
		\end{minipage}
		
		\caption{Visualization of cross-spectral density (CSD) matrix.}
		\label{fig3} 
	\end{figure}
	
	\subsection{Discussion}
	\subsubsection{The quality of generated signals}
	\begin{figure}[tb]
		\centering  %居中
		%\subfigure[Classic Cycle GAN transfer]{
			
			\includegraphics[width=\linewidth]{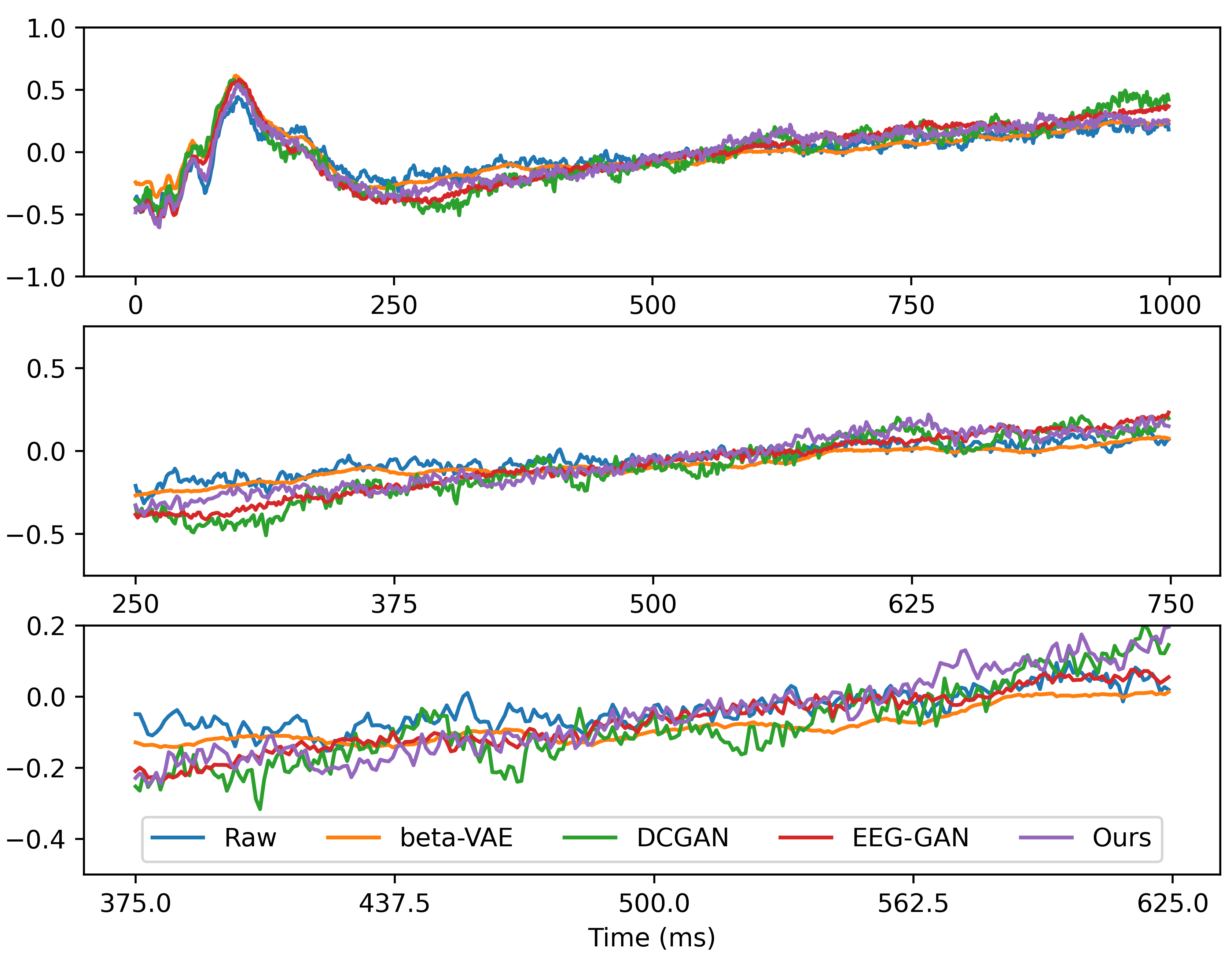}
			\caption{Comparison of generated signals by different methods.}
			\label{fig4} 
		\end{figure}
	
		In this section, we explore the relationship between augmentation size and classification accuracy. As shown in Table~\ref{tab2}, we manually set 5 groups of augmentation ratio, which are 1:10, 1:20, 1:30, 1:40, 1:50, respectively, and the results indicates that, bigger augmentation size does help for the generalization, therefore, improve performance. However, it saturates when reaching a certain threshold. It might because the augmented data can no longer introduce additional information to the target data distribution, (which also proves that information redundancy exists in the dataset). The topographic map (Fig.~\ref{erds}) proves our hypothesis in another way. By comparing between raw data and generated ones, we first observe a growing resemblance and after the data size increases to a certain degree, it almost imitates the unique characteristics of the target distribution. If keeping training in this phase, the uniqueness of augmented data will be compromised. Nonetheless, we state that the raw data is still of critical role in augmentation settings.
		
	\label{diss}
\subsubsection{The impact of augmentation size}
\begin{table}[tb]
	\caption{Influence of augmentation size.}
	\centering
	\resizebox{\linewidth}{!}{
		\begin{tabular}{c|ccccc}
			\toprule
			Baseline & x10 & x20 & x30& x40 &  x50\\ \midrule
			\specialrule{0em}{1pt}{1pt}
			79.95$\pm$9.91&81.78$\pm$9.34&\textbf{82.34$\pm$8.98}& 82.03$\pm$9.53& 81.09$\pm$9.85& 81.33$\pm$9.33\\
			
			\bottomrule
	\end{tabular}}
	\label{tab2}
\end{table}
		To validate the proposed method's effectiveness as a data augmentation approach, in Fig.~\ref{fig4}, we plot the averaged trials for different methods. It is remarkable that, our proposed method brings more detailed information especially when comparing with VAE, and it provides greater varieties compared with simply adding noise. It depicts the target distribution's flow and also introduce differences, hence, improve the generalization ability. The covariance matrix (Fig.~\ref{fig3}) suggests similar conclusion that both generalization and specificity are well considered in this approach. 
		
		\iffalse
		$$
		X_{n\times p}=\begin{pmatrix}
			x_{11}&\cdots&x_{1p} \\
			\vdots&\ddots&\vdots \\
			x_{n1}&\cdots&x_{np} \\
		\end{pmatrix}
		\stackrel{\text{PCA}}{\longrightarrow} \hat{X}_{n\times q}
		$$
		\fi
	
		\section{Conclusion and future works}
		\label{sec:conclusion}
		
		In this paper, we propose a novel data augmentation approach by transferring source data into target domain. The proposed method jointly consider the deficiencies of previous works on EEG data augmentation and the feasibility for further analysis on the target domain. For future work, we will explore the possibility of class-specific data transfer, which might provide target domain better disentanglement once upon the complete of data transfer.

\section*{Acknowledgment}
The authors thank to H.-B. Shin for useful discussions of the methods and experiments.

\bibliographystyle{IEEEtran}
\bibliography{ref}

\end{document}